\documentclass[aps,prl,twocolumn,twoside,superscriptaddress,nofootinbib,preprintnumbers]{revtex4-2}

\usepackage[bookmarks=false]{hyperref}
\usepackage{graphicx}
\usepackage{amsmath}
\usepackage{mathrsfs}
\usepackage{xspace}
\usepackage{xcolor}
\usepackage[normalem]{ulem}
\usepackage[abs]{overpic}
\usepackage{comment}
\usepackage{dsfont}
\usepackage{hyperref}
\usepackage{cleveref}
\usepackage{cancel}
\usepackage{amssymb}
\usepackage[normalem]{ulem}

\DeclareRobustCommand\wlout{\bgroup\markoverwith{\color{magenta}{\rule[0.4ex]{2pt}{0.8pt}}}\ULon}

\def\d{\mathrm{d}}
\newcommand{\vect}[1]{\boldsymbol{#1}}

\definecolor{hardcol}{RGB}{239, 187, 47}
\definecolor{beamcol}{RGB}{128, 0, 128}
\definecolor{jetcol}{RGB}{0, 0, 122}
\definecolor{softcol}{RGB}{255, 26, 255}

\allowdisplaybreaks[2]

\begin{document}


\title{Accessing nucleon transversity with one-point energy correlators}

\author{Mei-Sen Gao}
\email{msgao@fudan.edu.cn}
\affiliation{Department of Physics and Center for Field Theory and Particle Physics, Fudan University, Shanghai, 200433, China}

\author{Zhong-Bo Kang}
\email{zkang@physics.ucla.edu}
\affiliation{Department of Physics and Astronomy, University of California, Los Angeles, CA 90095, USA}
\affiliation{Mani L. Bhaumik Institute for Theoretical Physics, University of California, Los Angeles, CA 90095, USA}
\affiliation{Center for Frontiers in Nuclear Science, Stony Brook University, Stony Brook, NY 11794, USA}

\author{Wanchen Li}
\email{wanchenli@fudan.edu.cn}
\affiliation{Department of Physics and Center for Field Theory and Particle Physics, Fudan University, Shanghai, 200433, China}

\author{Ding Yu Shao}
\email{dyshao@fudan.edu.cn}
\affiliation{Department of Physics and Center for Field Theory and Particle Physics, Fudan University, Shanghai, 200433, China}
\affiliation{Key Laboratory of Nuclear Physics and Ion-beam Application (MOE), Fudan University, Shanghai, 200433, China}
\affiliation{Shanghai Research Center for Theoretical Nuclear Physics, NSFC and Fudan University, Shanghai 200438, China}
\affiliation{Center for High Energy Physics, Peking University, Beijing 100871, China}

\begin{abstract}
We propose a novel probe of the nucleon's transversity distribution, $h_1^q$, using the one-point energy correlator (OPEC), an infrared-and-collinear safe jet substructure observable. We demonstrate that in transversely polarized $p^{\uparrow}p$ collisions, the OPEC exhibits a single-spin asymmetry (SSA) with a clean $\sin(\phi_s - \phi_n)$ angular dependence. This method probes SSA over a much wider kinematic range in the angular scale $\theta_n$ compared to traditional measurements of hadron transverse momentum~$j_\perp$, establishing a complementary and systematically distinct channel to study the nucleon's three-dimensional structure at RHIC and the future Electron-Ion Collider.
\end{abstract}

\maketitle

{\it Introduction --}
A complete description of nucleon tomography is a central goal of hadron physics. Within the framework of collinear factorization, this structure at leading twist is completely characterized by three parton distribution functions (PDFs): the unpolarized ($f_1^q$), helicity ($g_1^q$), and transversity ($h_1^q$) distributions \cite{Ralston:1979ys, Jaffe:1991kp, Barone:2001sp}. While $f_1^q$ and $g_1^q$ are well constrained~\cite{Lin:2017snn}, the transversity PDF remains poorly known because of its unique chiral-odd nature \cite{Jaffe:1991kp}, which makes it inaccessible in inclusive scattering. Overcoming this experimental challenge is critical, as the first moment of $h_1^q$ defines the nucleon's tensor charge, $\delta q \equiv \int_0^1 dx \, [h_1^q(x) - h_1^{\bar{q}}(x)]$. This quantity not only provides deep insight into the nucleon spin structure, but also serves as a crucial input for BSM (Beyond the Standard Model) probes—such as in searches for new tensor interactions in neutron $\beta$-decay~\cite{Jackson:1957zz}. 

Phenomenologically, the transversity PDF has been accessed via single-spin asymmetries in semi-inclusive deep-inelastic scattering (SIDIS) and proton-proton collisions, primarily through the Collins effect~\cite{Collins:1992kk, Collins:1993kq, Kang:2015msa, Cammarota:2020qcw, Gamberg:2022kdb} and dihadron production channels~\cite{Radici:2018iag, Cocuzza:2023oam, Cocuzza:2023vqs}. In parallel with these efforts, direct Lattice QCD simulations have achieved remarkable precision~\cite{Gao:2023ktu}, enabling innovative analyses that use the lattice results as a direct constraint on the phenomenological fits~\cite{Lin:2017stx, Gamberg:2022kdb, Cocuzza:2023oam, Cocuzza:2023vqs} to help resolve historical tensions; see also recent developments in~\cite{Cammarota:2020qcw, Gamberg:2022kdb, Cocuzza:2023oam, DAlesio:2020vtw, Benel:2019mcq, Radici:2018iag}. Despite this impressive progress, current extractions from experimental data are limited by their reliance on non-perturbative fragmentation functions (FFs), which carry significant modeling uncertainties.

A promising alternative lies in the field of jet substructure. The energy–energy correlator (EEC)~\cite{Basham:1978bw, Basham:1978zq}, an infrared and collinear (IRC) safe event shape observable~\cite{Kinoshita:1962ur, Lee:1964is}, provides direct access to jet substructure. EEC has been extensively measured in $e^+e^-$ annihilation~\cite{SLD:1994idb, L3:1992btq, OPAL:1991uui, TOPAZ:1989yod, TASSO:1987mcs, JADE:1984taa, Fernandez:1984db, Wood:1987uf, CELLO:1982rca, PLUTO:1985yzc, OPAL:1990reb, ALEPH:1990vew, L3:1991qlf, SLD:1994yoe} and at the LHC~\cite{ATLAS:2015yaa, ATLAS:2017qir, ATLAS:2020mee, ALICE:2024dfl, CMS:2024mlf, CMS:2025ydi, ALICE:2025igw, ATLAS:2023tgo,CMS:2025jam}. Precision studies have also been proposed for SIDIS~\cite{Li:2021txc, Neill:2022lqx, Kang:2023oqj} at the future Electron–Ion Collider (EIC)~\cite{Accardi:2012qut, AbdulKhalek:2021gbh}. Recently, there has been growing interest in employing EEC to explore the nucleon tomography~\cite{Liu:2022wop, Liu:2023aqb, Cao:2023oef, Li:2023gkh, Kang:2023big, Bhattacharya:2025bqa, Kang:2024dja, Cuerpo:2025zde, Song:2025bdj}. For a recent review on energy correlators, see Ref.~\cite{Moult:2025nhu}.

In this Letter, we propose that one simple jet substructure observable—the one-point energy correlator (OPEC) \cite{Basham:1978bw, Basham:1978zq}—provides a novel and direct probe of the transversity PDF. This observable measures the angular distribution of energy flux, $\left\langle\mathcal{E}(\hat{\vect{n}})\right\rangle$. The energy flow operator $\mathcal{E}(\hat{\vect{n}})$
\begin{equation}
    \mathcal{E}(\hat{\vect{n}})=\int_0^\infty \d t\lim_{r\to \infty} r^2 n^i T_{0i}(t,r \hat{\vect{n}}),
\end{equation}
deposits the energy of final-state particles in a specific direction $\hat{\vect{n}}$. While OPEC has garnered recent attention~\cite{Riembau:2024tom, Mi:2025abd, Riembau:2025wjc, Zhu:2025qkx}, its connection to spin physics, particularly to the transversity PDF, has remained unexplored.

\begin{figure}[t]
    \centering
    \vspace{1cm}
    \includegraphics[width=1\linewidth]{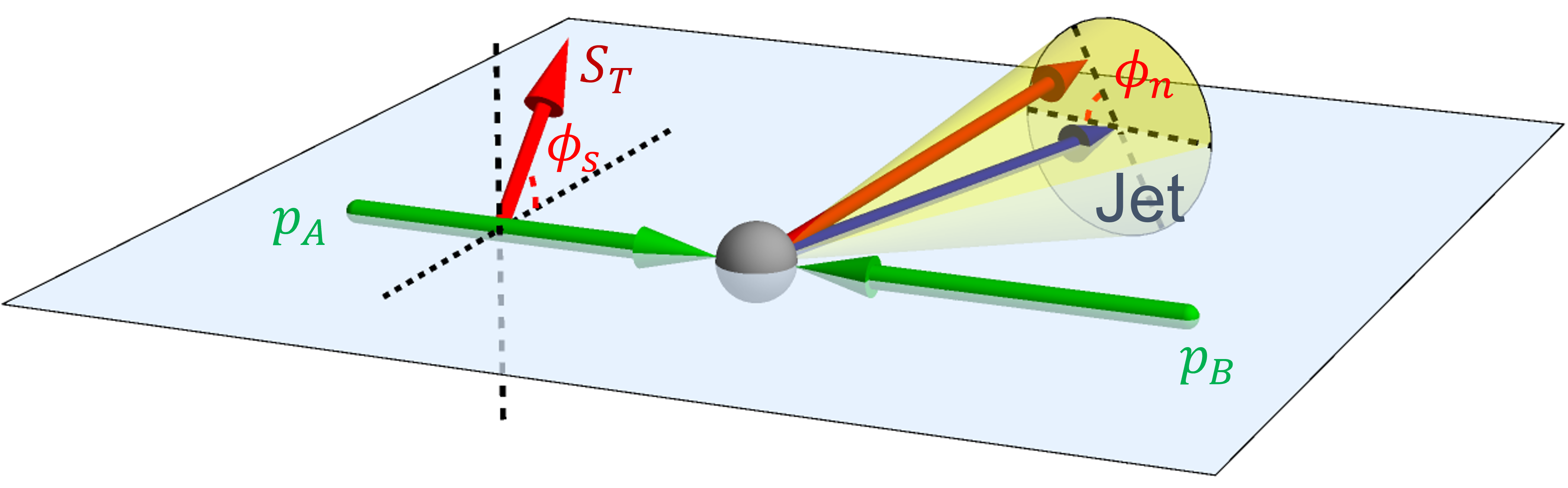}
    \caption{Kinematics of hadron production inside a jet in transversely polarized $p^{\uparrow}p$ collisions. The reaction plane is defined by the incoming proton momenta $p_A$, $p_B$, and the jet axis. The azimuthal angles $\phi_s$ and $\phi_n$ specify the orientations of the proton transverse spin vector $S_T$ and the energy detector with respect to this plane.}
    \label{fig:pp_collision}
\end{figure}

We demonstrate that in transversely polarized $p^{\uparrow}p$ collisions, this observable exhibits a clean single-spin asymmetry (SSA) with a characteristic $\sin(\phi_s - \phi_n)$ angular modulation, where $\phi_n$ and $\phi_s$ are the azimuthal angles of the energy flux detector and the nucleon spin relative to the reaction plane spanned by the incoming protons and the jet axis, as shown in Fig.~\ref{fig:pp_collision}. This measurement provides a new channel to access nucleon transversity with different dynamical weighting than existing methods. Compared with unweighted transverse momentum dependent (TMD) \cite{Boussarie:2023izj} studies \cite{Yuan:2007nd, D'Alesio:2011mc, Kang:2015msa, DAlesio:2017bvu, Kang:2017btw, Kang:2017glf},  the OPEC provides a cleaner input from a broader range of the jet substructure kinematics, while reducing the model dependence of non-perturbative fragmentation. This in turn enables a more robust determination of the transversity PDF. It offers a crucial, independent test of the universality of spin-dependent hadronization and establishes a powerful connection between the formal theory of energy flow and the non-perturbative partonic structure of the nucleon. In particular, this observable provides a timely avenue for phenomenological studies using Solenoidal Tracker at RHIC (STAR)  data~\cite{STAR:2017akg, STAR:2022hqg, Zhang:2024zuq, STAR:2025xyp}, and more broadly, this work opens a new frontier in the study of spin-dependent jet substructure and establishes a synergistic path between the ongoing RHIC spin program \cite{Nakagawa:2024rtw} and the precision measurements envisioned at the future EIC~\cite{Accardi:2012qut, AbdulKhalek:2021gbh}.

{\it OPEC. --}
In the collinear limit,  the OPEC is encoded within general light cone correlators that define a new class of fragmenting jet functions (FJFs) \cite{Procura:2009vm, Kang:2016ehg}. For a jet initiated by a quark, this correlator is given by
\begin{align}
\Delta^q(z, \hat{\vect{n}})
=\sum_X & \sum_{i \in J} \langle\Omega| \bar{\chi}_n \delta_{Q, \mathcal{P}_n} \delta^{(2)}(\hat{\vect{n}}-\hat{\vect{n}}_i)|J(h_i) X\rangle \nonumber \\
 & \times \frac{E_i}{E_J}\langle J(h_i) X| \chi_n|\Omega\rangle, 
\end{align}
where $\chi_{n}$ is the gauge invariant quark field in soft-collinear effective theory \cite{Bauer:2000ew, Bauer:2000yr, Bauer:2001ct, Bauer:2001yt} that creates collinear particles from the vacuum. $E_J$ and $E_i$ represent the energy of the jet and the measured hadron, respectively. $z\equiv E_J/Q$ represents the energy fraction of the jet that is initiated from the quark with energy $Q$. The measurement is performed on the energy flux inside the jet $J$, and the state $|J(h_i) X\rangle \notag$ represents the final-state unobserved particles $X$ and the jet $J$ with the hadron $h_i$ inside. This definition implicitly depends on the jet algorithms (e.g., the jet radius $R$), which we suppress for notational simplicity.  Taking into account the unpolarized and transversely polarized quark initiating the jet, we can parametrize the above correlators at the leading power as \cite{Boussarie:2023izj}  
\begin{equation}
    \Delta^{[\gamma^+]}_q  =\mathcal{J}^q,  \qquad \Delta^{[i\sigma^{\alpha +}\gamma_5]}_q  =\epsilon_T^{\alpha \beta}\hat{\vect{n}}_{T,\,\beta}\frac{p_J^-}{2}\theta_n \mathcal{J}^q_{1, \,\perp},
\end{equation}
where we have $p_J^- =\bar n\cdot p_J$, $\sigma^{\mu \nu} = i[\gamma^\mu,\gamma^\nu]/2$, $n\cdot \bar n =2$, and $\epsilon_T^{\alpha \beta} = \epsilon^{\alpha \beta \rho \sigma} \bar n_\rho n_\sigma/(n\cdot \bar n)$. $\hat{\vect{n}}_{T}$ is the transverse direction of the energy flow relative to the jet axis. $\mathcal{J}^q$ is the unpolarized OPEC FJF, while $\mathcal{J}_{1,\perp}^q$ denotes the transversely polarized OPEC FJF, which captures the fragmentation of a transversely polarized quark into an azimuthally asymmetric energy flux within the jet.

{\it Factorization formula. --}
We consider the OPEC in inclusive jet production for the polarized proton and unpolarized proton scattering process, $p^{\uparrow}+p\to J+X$. The OPEC observable is defined as
\begin{equation}\label{eq:definition}
    \begin{aligned}
    \!\frac{\d \Sigma}{\d \theta_n \d \phi_n \d \eta \, \d p_T} = &\sum_{h\in J}\int_0^1 \d z_h \int \d^2 \Omega_{h}\,  \delta\left(\phi_n -\phi_h \right)  \\
    &\times \delta\left(\theta_n-\theta_h\right) \, z_h\,\frac{\d\sigma}{\d z_h \, \d^2 \Omega_{h}\d \eta \, \d p_T}\,,
\end{aligned}
\end{equation}
where the jet is characterized by rapidity $\eta$ and transverse momentum $p_T$, and the sum runs over hadrons inside the jet. For each hadron $h$, the solid angle 
$\d^2 \Omega_{h} \equiv \sin\theta_h \, \d \theta_h \, \d \phi_h$ specifies its direction relative to the jet axis, with $\theta_h$ denoting the polar angle and $\phi_h$ the azimuthal angle. The weight factor $z_h$ represents the energy fraction of the fragmenting quark jet carried by the hadron.

The OPEC observable offers two key advantages over traditional hadron-in-jet distributions. First, the built-in integral over $z_h$ in \cref{eq:definition} projects the full $z_h$-dependence of the hadron distribution onto a single moment. As a result, a TMD fit involving OPEC requires only a parametrization in the Fourier variable $b$, rather than in the two-dimensional space $(b, z_h)$, substantially reducing the model dependence associated with final-state fragmentation~\cite{Kang:2024dja,Cuerpo:2025zde}. Second, OPEC benefits from angular measurements at far higher precision than hadron-in-jet studies~\cite{STAR:2017akg, STAR:2022hqg, Zhang:2024zuq, STAR:2025xyp, Yuan:2007nd, D'Alesio:2011mc, Kang:2015msa, DAlesio:2017bvu, Kang:2017btw, Kang:2017glf}, where substructure is accessed through the hadron’s transverse momentum $j_\perp\simeq p_T z_h \theta_n$ in the collinear limit. For STAR kinematics at $\sqrt{s}=510\,\text{GeV}$ and $p_T=32.3\,\text{GeV}$ with a typical cut $z_h=0.1$, the accessible $j_\perp$ range $(0.1,1)$ GeV corresponds to $\theta_n\simeq(0.03,0.3)$ rad. By contrast, modern detectors achieve resolutions of order $10^{-3}$ rad: CMS has measured energy correlators below this scale~\cite{CMS:2024mlf}, while the EIC is expected to reach down to $10^{-4}$ rad for both polar and azimuthal angles at high $p_T$~\cite{Arrington:2021yeb, AbdulKhalek:2021gbh, Milton:2024bqv}. The OPEC framework thus enables jet substructure studies at angular scales more than an order of magnitude finer than previously probed, while remaining IRC safe and robust in the small-angle regime.

The azimuthal dependence of the OPEC distribution can be decomposed as
\begin{equation}
    \frac{\d\Sigma}{\d \theta_n \d \phi_n \d \eta \, \d p_T}  =  Z_{UU} + \sin{(\phi_s-\phi_n)}Z_{UT},
\end{equation}
where the unpolarized ($Z_{UU}$) and spin-dependent ($Z_{UT}$) structure functions take the factorized forms:
\begin{align}
& Z_{UU}  = \frac{\alpha_s^2}{s} p_T^2 \theta_n\sum_{a,b,c} \int \frac{\d x_1}{x_1} f_{a/A}(x_1, \mu) \int \frac{\d x_2}{x_2} f_{b/B}(x_2, \mu) \nonumber \\
&\hspace{0.5cm}\times \mathcal{J}^{c}(\theta_n,Q) H_{ab\to c}^{\rm U}(\hat s, \hat t, \hat u) \delta(\hat s + \hat t + \hat u)\,,
\label{eq:ZUU}
\\
& Z_{UT}  = \frac{\alpha_s^2}{s}p_T^2 \theta_n \sum_{a,b,c} \int \frac{dx_1}{x_1} h_1^{a}(x_1, \mu)  \int \frac{dx_2}{x_2} f_{b/B}(x_2, \mu) \nonumber \\
 &\hspace{0.5cm} \times p_T \theta_n\mathcal{J}^{c}_{1,\perp}(\theta_n,Q) H_{ab\to c}^{\rm Collins}(\hat s, \hat t, \hat u) \delta(\hat s + \hat t + \hat u)\,.
\label{eq:ZUT}
\end{align}
Here, $\alpha_s$ denotes the strong coupling constant and $s$ is the square of the center-of-mass energy of the proton-proton system. The hard function $H_{ab\to c}^{\rm U}$ and $H_{ab\to c}^{\rm Collins}$ represent the unpolarized and spin-dependent partonic matrix elements, respectively, for the subprocess $ab \to c$~\cite{Owens:1986mp, Yuan:2007nd, Kang:2017btw, Kang:2010zzb}. The variables $\hat s$, $\hat t$, and $\hat u$ are the Mandelstam variables of the partonic interaction. The collinear PDF $f_a$ and transversity PDF $h_1^a$ depend on the factorization scale $\mu$, while the energy-weighted jet functions $\mathcal{J}^{c}$ and $\mathcal{J}_{1,\perp}^{c}$ are evaluated at the hard scale $Q$.

{\it Relations to the Collins function. --}
A powerful application of our formalism is to perform direct tests of the universality of polarized fragmentation. The Collins function~\cite{Collins:1992kk, Collins:1993kq}, which encodes the information of the azimuthal distribution of hadrons fragmenting from a transversely polarized quark, has been a cornerstone for accessing transversity, with its universality across different scattering processes being a crucial, yet challenging, assumption to verify \cite{Boer:2003cm, Metz:2002iz, Collins:2004nx, Yuan:2009dw, Boer:2010ya}.

Our polarized FJF, $\mathcal{J}_{1,\perp}^{q}$, provides a direct probe of the same underlying physics but in terms of energy flow rather than a specific hadron species. This distinction is paramount. As an IRC safe observable, the OPEC FJF is theoretically cleaner and less susceptible to the complexities of hadronization models. Universality can therefore be tested by comparing measurements of $\mathcal{J}_{1,\perp}^{q}$ in distinctly different environments: (1) $e^+e^-$ annihilation, the cleanest environment where jets are produced from a quark-antiquark pair; (2) SIDIS, probing the fragmentation of a quark struck from within a nucleon; and (3) polarized proton-proton collisions, the most complex hadronic environment. A comparison of the extracted $\mathcal J_{1,\perp}^{q}$ from these channels, after accounting for their respective QCD evolution, would constitute a rigorous and independent test of the factorization and universality of the polarized fragmentation mechanism. 

In the limit $\theta_n\gg\Lambda_{\rm QCD}/Q$, the unpolarized and transversely polarized OPEC FJFs can be systematically matched onto their collinear counterparts using the operator product expansion (OPE)
\begin{align}\label{eq:FJF_ope}
    &\mathcal{J}^{q}(\theta_n,Q) =\sum_h \int_0^1 \d z_h z_h \int_0^{\infty} \frac{\d b\, b}{2\pi} J_0\left(p_T \theta_n b\right)\nonumber\\
    & \hspace{1.2cm}\times  \hat C_{i\gets q}^D\otimes D_{h/i}(z_h, \mu_b)  e^{-\frac{1}{2} S_{\rm pert}(Q, b)}  ,  \\
&p_T \theta_n\mathcal{J}^{q}_{1,\perp}(\theta_n,Q)  = \sum_h \int_0^1 \d z_h z_h\int_0^{\infty} \frac{\d b\, b^2}{2\pi} J_1\left(p_T \theta_n b\right) \nonumber \\
&\hspace{1.2cm}\times \delta \hat C_{i\gets q}^{\rm Collins} \otimes \hat H^{(3)}_{h/i}(z_h, \mu_b)  e^{-\frac{1}{2} S_{\rm pert}(Q, b)},
\label{eq:FJF_ope_Collins}
\end{align}
where we have included QCD evolution in the perturbative Sudakov factor $S_{\rm pert}$ \cite{Boussarie:2023izj}.  Here $J_0$ and $J_1$ are Bessel functions,  $\mu_b = 2 e^{-\gamma_E}/b$ is the  OPE scale of the collinear FFs. $D_{h/i}$ is the unpolarized collinear FF, and $\hat H_{h/i}^{(3)}$ is the first moment of the Collins function in the convention of Ref.~\cite{Yuan:2009dw, Kang:2015msa}. In the expressions above, we neglect the resummation of jet radius logarithms, and the complete factorized structure and renormalization group evolution of the OPEC FJFs are detailed in the supplemental material. The usual convolution $\otimes$ is defined as
\begin{align}
    \hat C_{i\gets q}\otimes F_{h/i}
=& \sum_{i} \int_{z_h}^1\frac{\d \xi}{\xi}   \hat C_{i\gets q}\left(\xi, \mu_b\right)F_{h/i}\left(\frac{z_h}{\xi}, \mu_b\right),
\end{align}
where $F\in\{D,\, \hat H^{(3)}\}$ and we have ignored the two-variable twist-$3$ FFs that appear in the collinear factorization region~\cite{Yuan:2009dw}. At LO, both matching coefficients $\hat C_{i\gets q}^{D}$ and $\delta \hat C_{i\gets q}^{\rm Collins}$ reduce to $\delta_{qi}\, \delta(1 - \xi)$, while beyond LO, the matching coefficients generally depend on the jet algorithm~\cite{Kang:2017glf, Kang:2020xyq, Kang:2023elg, Generet:2025vth}. The complete one-loop expressions for the anti-$k_T$ algorithm \cite{Cacciari:2008gp} are provided in the supplemental material.

{\it Single spin asymmetry. --}
The transverse single-spin asymmetry (SSA) of the OPEC is given by
\begin{equation}
 A_{UT}^{\sin(\phi_s - \phi_n)} = \frac{Z_{UT}}{Z_{UU}}.
\end{equation}
In our numerical calculations, we have included contributions from the gluon-initiated jet channel, which enter the unpolarized structure function $Z_{UU}$ and thus influence the normalization of the SSA. We perform numerical evaluations at next-to-leading logarithmic (NLL) accuracy, setting $\mu = Q = p_T$ and employing the LO matching coefficients in \cref{eq:FJF_ope,eq:FJF_ope_Collins}. The perturbative Sudakov factors are detailed in the supplemental material. A more complete treatment, including the resummation of jet radius, which is necessary for achieving higher theoretical precision, will be presented in a future work. In addition, the full Sudakov exponential is modeled as $e^{-S_{\text{pert}}(Q, b_*)/2 - S_{\rm NP}(Q, b)}$. In $S_{\text{pert}}$, we separate perturbative and non-perturbative domains using the standard $b_*$-prescription, where $b_* = b/\sqrt{1+b^2/b_{\rm max}^2}$ freezes the impact parameter at $b_{\rm max} = 1.5~\text{GeV}^{-1}$~\cite{Collins:1984kg, Qiu:2000ga, Landry:2002ix}. This prevents the evolution from entering a region where perturbation theory is unreliable. For the unpolarized FJF, the non-perturbative component $S_{\rm NP}^{D_1}(Q, b)$ is adapted from the well-established form used for Drell-Yan and SIDIS processes~\cite{Sun:2014dqm}. For the transversely polarized FJF, the corresponding non-perturbative factor $S_{\rm NP}^{\rm Collins}(Q, b)$ is taken from global analyses of the Collins effect~\cite{Kang:2015msa}.

\begin{figure}[t!]
    \centering
    \includegraphics[width=1\linewidth]{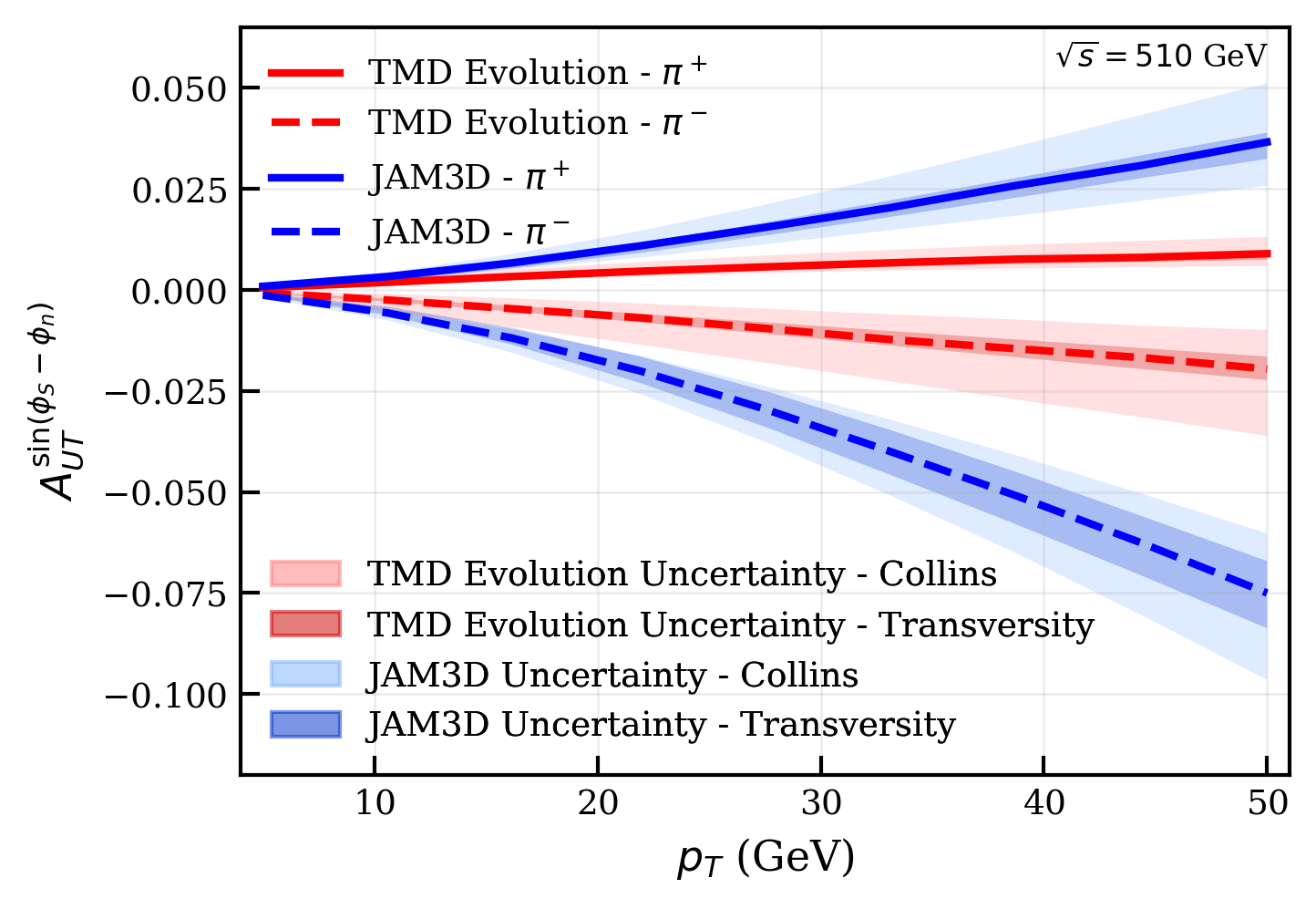}
    \caption{SSA $A_{UT}^{\sin(\phi_s - \phi_n)}$ as a function of jet transverse momentum $p_T$ for $\pi^{\pm}$ at $\sqrt{s} = 510\, {\rm GeV}$, evaluated using two approaches: the JAM3D-22 and the TMD evolution framework. Error bands represent uncertainties from the transversity PDF and the Collins FF parametrizations, respectively. The $p_T$ distribution is obtained by integrating over $\theta_n \in [0,0.1]$.}
    \label{fig: pT bands}
\end{figure}

Furthermore, since only charged $\pi^\pm$ Collins FFs are known, we apply a less inclusive version of our formula, replacing the sum over all hadrons $\sum_{h}$ in ~\cref{eq:FJF_ope,eq:FJF_ope_Collins} with a subset of hadrons sharing the same quantum numbers, $\sum_{h \in \mathbb{S}}$~\cite{Lee:2023tkr, Lee:2023npz, Li:2021zcf, Chen:2020vvp, Jaarsma:2023ell}. In this work, we only consider charged pions $\pi^\pm$ and present the SSAs $A_{UT}^{\sin(\phi_s - \phi_n)}$ in two approaches:
\begin{itemize}
    \item \textit{TMD evolution framework}: The full TMD evolution formalism given in~\cref{eq:FJF_ope,eq:FJF_ope_Collins} at NLL accuracy, where we parametrize the transversity PDF and Collins FF according to \cite{Kang:2015msa}, and retain the same collinear inputs as in that analysis: CT10nlo PDF \cite{Lai:2010vv} and NLO DEHSS FF~\cite{deFlorian:2014xna}.
    \item \textit{JAM3D framework}: The approach of the JAM3D-22 global analysis \cite{Gamberg:2022kdb}. Here, the Sudakov exponentiation (with Fourier transform) in Eqs.~(8) and (9) is replaced by a Gaussian modeling of the transverse momentum dependence, and the corresponding collinear components are evolved using  Dokshitzer-Gribov-Lipatov-Altarelli-Parisi equations.
\end{itemize}

In \cref{fig: pT bands}, we present SSA $A_{UT}^{\sin(\phi_s - \phi_n)}$ as a function of jet transverse momentum $p_T$ at $\sqrt{s} = 510~ {\rm GeV}$, computed using both the JAM3D-22 and full evolution frameworks. The evolved prediction that is based on the global fit of \cite{Kang:2015msa} yields systematically smaller asymmetries compared to the JAM3D-22 result. We also analyze the uncertainties originating from two key sources: the transversity PDFs and the Collins FFs. In both approaches, we observe that the dominant source of uncertainty arises from the Collins function, while the transversity contribution plays a comparatively smaller role.

\begin{figure}[t]
    \centering
    \includegraphics[width=0.95\linewidth]{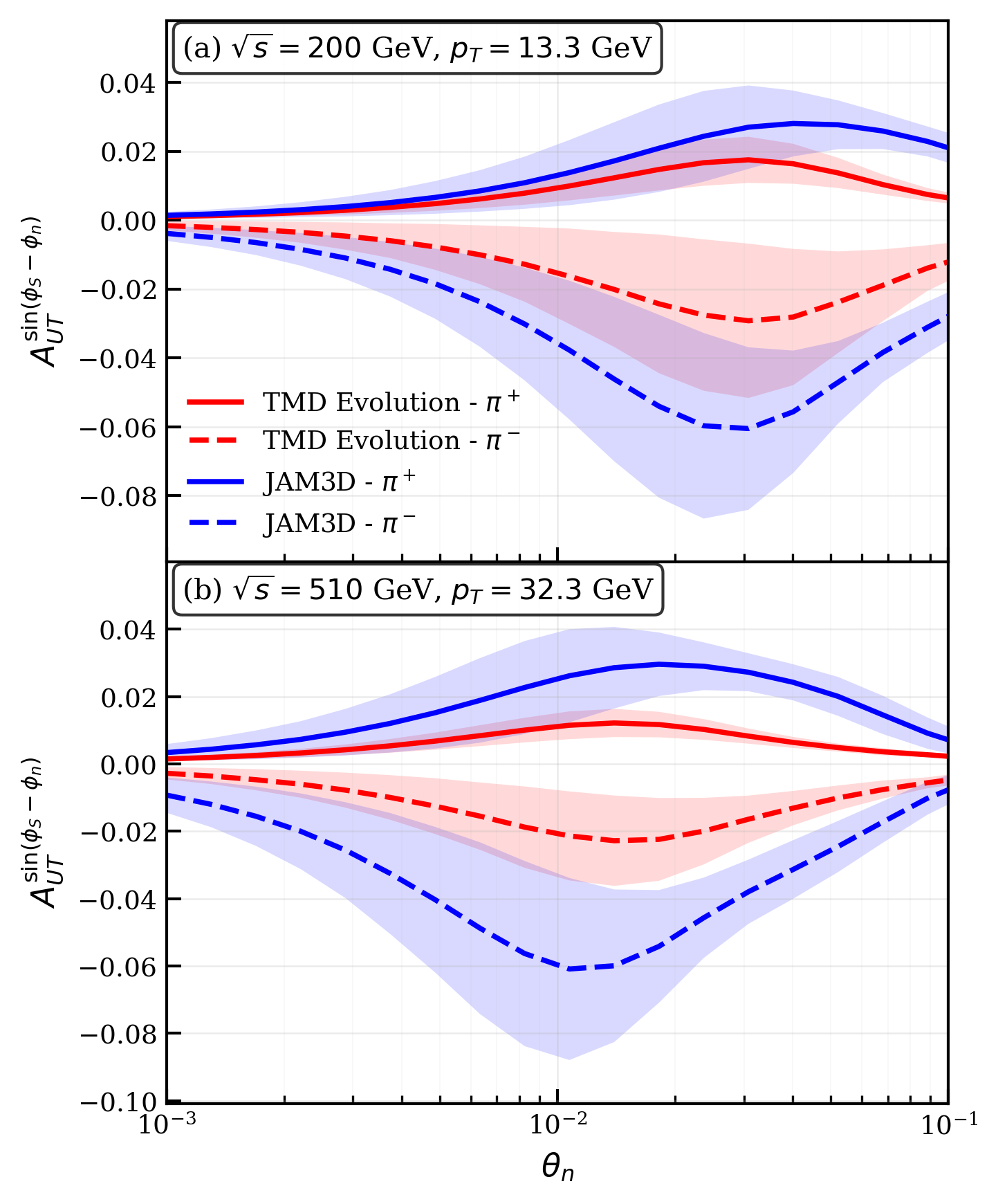}
    \caption{SSA $A_{UT}^{\sin(\phi_s - \phi_n)}$ as a function of $\theta_n$ at (a) $\sqrt{s} = 510\, \text{GeV}$, $p_T = 32.3\, \text{GeV}$, (b) $\sqrt{s} = 200\, \text{GeV}$, $p_T = 13.3\, \text{GeV}$, computed using the JAM3D-22 and TMD evolution frameworks.  Error bands represent combined uncertainties from transversity PDFs and Collins FFs.}
    \label{fig: thetaT_combined}
\end{figure}

In \cref{fig: thetaT_combined}, we show the SSA $A_{UT}^{\sin(\phi_s - \phi_n)}$ as a function of the angle $\theta_n$ of the energy detectors at two kinematic settings extensively studied by the STAR Collaboration~\cite{STAR:2017akg, STAR:2022hqg, Zhang:2024zuq, STAR:2025xyp}: (a) $\sqrt{s} = 510\, \text{GeV}$, $p_T = 32.3\, \text{GeV}$, (b) $\sqrt{s} = 200\, \text{GeV}$, $p_T = 13.3\, \text{GeV}$, respectively. The error bands reflect combined uncertainties from the transversity PDFs and Collins FFs. The energy-weighted asymmetries exhibit similar magnitudes across the two collision energies, consistent with recent observations of energy independence of the SSA in $p^\uparrow p$ collisions \cite{STAR:2025xyp}. However, their peak positions in $\theta_n$ differ, reflecting a shift in the angular structure of the fragmentation dynamics. Moreover, the partial overlap between the error bands from the TMD evolution and JAM3D approaches suggests that current parametrizations do not yet allow us to resolve the effects of TMD evolution. Nonetheless, the reduced overlap observed at higher $p_T$ highlights the potential for future data to provide clearer resolution.

\begin{figure}[t]
    \centering
    \includegraphics[width=1\linewidth]{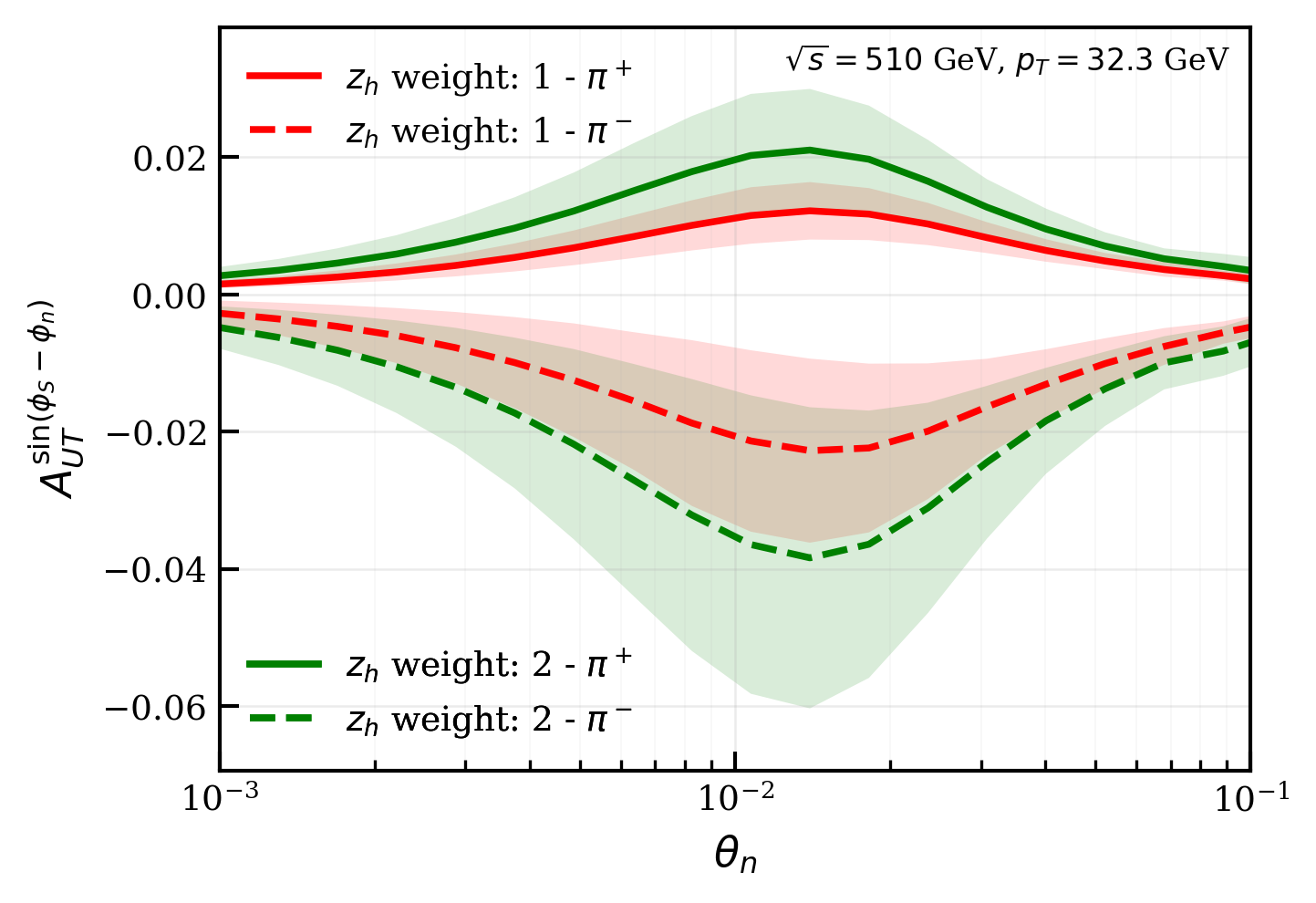}
    \caption{SSA $A_{UT}^{\sin(\phi_s - \phi_n)}$ as a function of $\theta_n$ at $\sqrt{s} = 510\, {\rm GeV}$ and fixed jet transverse momentum $p_T = 32.3\, {\rm GeV}$, computed for two different $z_h^{(n)}$ weights $n =1,\,2$ in TMD evolution framework. The error bars reflect combined uncertainties from both the transversity parton distribution functions and the Collins FF parametrizations.}
    \label{fig: thetaT zweight}
\end{figure}

{\it Generalization. --}
In the standard definition of OPEC in \cref{eq:definition}, the energy weight $z_h$ enters linearly. A natural generalization is to raise $z_h$ to an arbitrary power, thereby probing higher Mellin moments of the Collins FFs.
In \cref{fig: thetaT zweight}, we compare the resulting SSA as functions of $\theta_n$ for different choices of the $z_h$ weight.
We observe that increasing the $z_h$ weight, i.e. higher Mellin moment, leads to a remarkable enhancement in the magnitude of the SSA.

{\it Conclusions and outlook. --}
We introduced the OPEC in jets for transversely polarized $p^{\uparrow}p$ collisions, sensitive to the transverse polarization effect through a $\sin(\phi_s -\phi_n)$ modulation. Defined via an energy-weighted angular distribution of energy flux detectors inside jets, this observable provides a clean probe of the transversity PDF and Collins FF at high momentum scales. Compared with prior studies, we estimate that OPEC can access a much broader kinematic range of the jet substructure.
 
We also present a phenomenology study of SSA from OPEC via two frameworks: a) a TMD evolution framework using global fits from \cite{Kang:2015msa}, b) the JAM3D-22 global analysis \cite{Gamberg:2022kdb}, which omits TMD evolution. Both approaches reveal similar qualitative behavior, with partially overlapping uncertainty bands between the two approaches. The OPEC SSA offers a complementary channel to constrain the transversity PDF and Collins function at RHIC. Looking forward, implementing this observable at the future EIC would enable a pristine test of the fundamental principle of universality for the Collins FF by allowing a direct comparison between hadronic ($p^{\uparrow}p$) and deep-inelastic scattering ($e p^{\uparrow}$) processes. This establishes a vital new tool for precision studies of spin-dependent fragmentation and the rich internal landscape of the nucleon.

{\it Acknowledgements --}
We thank Yiyu Zhou for helpful discussions. M.G., W.L., and D.Y.S. are supported by the National Science Foundations of China under Grant No.~12275052, No.~12147101, No.~12547102. D.Y.S. is also supported by the Innovation Program for Quantum Science and Technology under grant No. 2024ZD0300101. Z.K. is supported by the National Science Foundation under grant No.~PHY-2515057.

\bibliography{refs.bib}

\clearpage
\appendix
\onecolumngrid
\setcounter{equation}{0}
\renewcommand{\theequation}{S-\arabic{equation}}
\setcounter{figure}{0}
\renewcommand{\thefigure}{S-\arabic{figure}}
 \allowdisplaybreaks
\section*{Supplemental Material} 

\subsection*{One-point energy correlator fragmenting jet functions}
This section provides additional details on the one-point energy correlator (OPEC) fragmenting jet functions (FJFs), including their factorization structure, matching onto the collinear fragmentation functions (FFs), and the evolution properties. Analogously to the TMD FJFs~\cite{Kang:2017glf,Kang:2023elg}, the OPEC FJFs admit the factorized representations in the limit $R\gg \theta_n$,
\begin{align} 
\label{eq: OPEC JU}
    \mathcal{J}^{c}(z, \hat{\vect{n}}, \omega_J R, \mu)  =&  \, \mathcal{H}_{c\to i}(z, \omega_J R, \mu) 
\int \d^2 {\vect l}_\perp \d^2{\vect \lambda}_\perp \delta^2\left({\vect l}_\perp + p_T\theta_n \hat{\vect{n}}_T-{\vect \lambda}_\perp \right)
\nonumber \\
&\times J_{i}^U( {\vect l}_\perp, \mu, \nu) S_i({\vect \lambda}_\perp, \mu, \nu R)\; ,\\
 p_T \theta_n\mathcal{J}^{q}_{1,\perp} (z,\hat{\vect{n}}, \omega_J R, \mu)  =& \, \mathcal{H}_{c\to i}^{T}(z, \omega_J R, \mu) 
\int \d^2 {\vect l}_\perp \d^2{\vect \lambda}_\perp \delta^2\left({\vect l}_\perp + p_T\theta_n \hat{\vect{n}}_T-{\vect \lambda}_\perp \right)
\nonumber \\
&\times  p_T\theta_n J^{T}_{i}({\vect l}_\perp, \mu, \nu) S_i({\vect \lambda}_\perp, \mu, \nu R)\;.
\label{eq: OPEC JT}
\end{align}
where $\omega_J = \bar n\cdot p_J$ and $R$ is the jet radius. The unit vector $\hat{\vect n}_T$ denotes the transverse direction of the energy flow with respect to the jet axis. $\vect \lambda_\perp$ is the transverse momentum of the soft radiation, and $\vect l_\perp$ is  the transverse momentum of the  fragmenting parton $i$ relative to the energy flow direction $\hat{\vect{n}}$.

In the above expressions, $\mathcal{H}_{c\to i}(z, \omega_J R, \mu)$ and $\mathcal{H}^{T}_{c\to i}(z, \omega_J R, \mu)$ are the unpolarized and transversely polarized hard matching coefficients. They encode the contributions from radiations at the jet scale $\omega_J R$, corresponding to energetic emissions that are close to the jet boundary.
For completeness, we collect the anti-$k_T$ results known up to next-to-leading order (NLO)~\cite{Kang:2017glf, Kang:2023elg}
\begin{align}
    \mathcal{H}_{q\to q'}(z,\omega_J R,\mu) 
  &= \delta_{qq'}\delta(1-z) +\delta_{qq'} \frac{\alpha_s}{2\pi} \bigg[C_F \delta(1-z)\Big(-\frac{L^2}{2} - \frac{3}{2} L +\frac{\pi^2}{12} \Big) + P_{qq}(z) L -2C_F(1+z^2)\left(\frac{\ln(1-z)}{1-z}\right)_+
\nonumber \\
 & \quad
 -C_F(1-z)  \bigg] \,, 
\\
 \mathcal{H}_{q\to g}(z,\omega_J R,\mu) 
 &=\frac{\alpha_s}{2\pi}\bigg[\Big(L - 2 \ln(1-z) \Big) P_{gq}(z) - C_Fz \bigg]\,, 
 \\
\mathcal{H}_{g\to g}(z, \omega_J R, \mu) 
& = \delta(1-z) + \frac{\alpha_s}{2\pi}\bigg[ \delta(1-z)\Big(-C_A\frac{L^2}{2} - \frac{\beta_0}{2} L + \frac{\pi^2}{12}\Big) + P_{gg}(z) L
 - \frac{4C_A (1-z+z^2)^2}{z} \left(\frac{\ln(1-z)}{1-z}\right)_{+} \bigg]\,,
\\
\mathcal{H}_{g\to q}(z,\omega_J R, \mu) 
 & =  \frac{\alpha_s}{2\pi}\bigg[\Big(L - 2\ln(1-z) \Big)  P_{qg}(z) - T_F 2z(1-z) \bigg]\,,
 \\
     \mathcal{H}^T_{q \to q'} (z, \omega_J R, \mu)
& =
\delta _{qq'} \delta(1-z)
+ \delta _{qq'} \frac{\alpha _s}{2 \pi} \bigg[C_F \delta(1-z) \left(-\frac{3}{2} L - \frac{L^2}{2} + \frac{\pi ^2}{12} \right)+ \Delta _T P_{qq}(z) L  -4 C_F z \left(\frac{\ln(1-z)}{1-z}\right)_+ \bigg].
\end{align}
with $L \equiv \ln\left[ { \mu^2} / \left({ \omega_J^2 \tan^2 (R/2)} \right)\right]$, $C_F =(N_C^2-1)/(2N_C)$, $C_A=N_C$, $T_F=1/2$, $\beta_0=(11C_A-4n_fT_F)/3$, $N_C$ the number of colors, and $n_f$ the number of quark flavors. The LO DGLAP splitting functions are
\begin{align}
    P_{qq}(z) &= C_F \left[\frac{1+z^2}{(1-z)_+} + \frac{3}{2}\delta(1-z) \right]\,,
\label{eq:Pqq}
\\
P_{gq}(z) &= C_F \frac{1+(1-z)^2}{z}\,,
\\
P_{gg}(z) &= 2C_A \left[\frac{z}{(1-z)_+} + \frac{1-z}{z} +z(1-z) \right] + \frac{\beta_0}{2} \delta(1-z)\,,
\label{eq:Pgg}
\\[.2cm]
P_{qg}(z) &= T_F\left[z^2+(1-z)^2\right]\,.
\\
\Delta _T P_{qq}(z)
& =
C_F \left[\frac{2z}{(1-z)_+} + \frac{3}{2} \delta (1-z)\right].
 \label{e.AP_transverse_qq}
\end{align}
The soft function $S_i(\vect{\lambda}_\perp, \mu, \nu R)$ captures soft radiation constrained within the jet boundary and is known up to NLO~\cite{Kang:2017glf}
\begin{align}
    S_i^{\rm bare}({\vect b}, \mu, \nu R) &= \int \d^2{\vect \lambda}_\perp e^{-i{\vect \lambda}_\perp\cdot {\vect b}} S_i({\vect \lambda}_\perp, \mu, \nu R) 
\label{eq:softb}\\
=  1 + \frac{\alpha_s}{2\pi} C_i &\bigg[ \frac{2}{\eta} \left( - \frac{1}{\epsilon}-\ln\left(\frac{\mu^2}{\mu_b^2}\right)\right) + \frac{1}{\epsilon^2} -\frac{1}{\epsilon} \ln\left(\frac{\nu^2 \tan^2 (R/2)}{\mu^2}\right)
 - \ln\left(\frac{\mu^2}{\mu_b^2}\right) \ln\left(\frac{\nu^2 \tan^2(R/2)}{\mu_b^2}\right) + \frac{1}{2}\ln^2\left(\frac{\mu^2}{\mu_b^2}\right) - \frac{\pi^2}{12} \bigg]. \nonumber
\end{align}
Here $\mu_b \equiv 2e^{-\gamma_E}/b$ and $C_i=C_F\, ( C_A)$ for $i=q\,(g)$. The renormalization of the soft function takes the form
\begin{equation}
    S_i(\vect b, \mu, \nu R) = Z^S_i({\vect b}, \mu, \nu)S^{\rm bare}_i (\vect b, \mu, \nu R),
\end{equation}
with 
\begin{equation}
    Z^S_i(b, \mu, \nu) = 1+\frac{\alpha_s C_i}{2\pi}\left[ \frac{2}{\eta} \left( - \frac{1}{\epsilon}-\ln\left(\frac{\mu^2}{\mu_b^2}\right)\right) + \frac{1}{\epsilon^2} -\frac{1}{\epsilon} \ln\left(\frac{\nu^2 \tan^2 (R/2)}{\mu^2}\right) \right].
\end{equation}

The functions $J_i^U$ and $J_i^T$ denote, respectively, the unsubtracted unpolarized and transversely polarized OPEC jet functions. The renormalization of the OPEC jet functions takes the form
\begin{align}
    J_{i}^{U}( \vect b, \mu, \nu)  &= Z_i^J(\vect b,\mu,\nu) J_{i}^{U\, \rm bare}( \vect b, \mu, \nu), 
    \label{eq:JUjet}\\
    J_{q}^{T}( \vect b, \mu, \nu) & = Z_q^J(\vect b,\mu,\nu) J_{q}^{T\, \rm bare}( \vect b, \mu, \nu),
    \label{eq:JTjet}
\end{align}
with $i=q,g$ and
\begin{align}
    Z_q^J(b,\mu,\nu)&= 1+\frac{\alpha_s C_F}{2\pi}\left[ \frac{2}{\eta}\left( \frac{1}{\epsilon}+\ln \frac{\mu^2}{\mu_b^2}\right)+\frac{1}{\epsilon}\left( 2\ln \frac{\nu}{\omega_J}+\frac{3}{2}\right)\right], 
    \label{eq:ZqJ}\\
    Z_g^J(b,\mu,\nu)&= 1+\frac{\alpha_s C_A}{2\pi}\left[ \frac{2}{\eta}\left( \frac{1}{\epsilon}+\ln \frac{\mu^2}{\mu_b^2}\right)+\frac{1}{\epsilon}\left( 2\ln \frac{\nu}{\omega_J}+\frac{\beta_0}{2C_A}\right)\right].
    \label{eq:ZgJ}
\end{align}
The rapidity divergence between the soft function in \cref{eq:softb} and OPEC jet functions cancels in \cref{eq:JUjet,eq:JTjet,eq:ZqJ,eq:ZgJ}. Consequently, the genuine OPEC jet functions can be defined as
\begin{equation}
    {\tilde{J}}_{i}^{U/T}(\vect b, \mu) \equiv J_i^{U/T}( \vect b, \mu, \nu)\, S_i(\vect b, \mu, \nu R)\,,
\end{equation}

We also compute the NLO perturbative matching of the $z_h^{N-1}$-weighted OPEC jet functions onto the corresponding partonic collinear FF counterparts in the limit $b^2\Lambda_{\rm QCD}^2\ll 1$
\begin{align}
    \label{eq: JU bare}
    \tilde{J}_{q}^{U}({\vect b}, \mu,\nu) &= C_{qq}(\vect b, \mu,\nu) \,\tilde D^{N}_{q/q}(\mu) + C_{gq}(\vect b, \mu,\nu) \,\tilde D^{N}_{g/q}(\mu), \\
    \label{eq: JUg bare}
    \tilde{J}_{g}^{U}({\vect b}, \mu,\nu) &= C_{gg}(\vect b, \mu,\nu) \,\tilde D^{N}_{g/g}(\mu) + C_{qg}(\vect b, \mu,\nu) \,\tilde D^{N}_{q/g}(\mu), \\
    \label{eq: JT bare}
    \tilde{J}_{q}^{T}({\vect b}, \mu,\nu) &= \delta C_{qq}( \vect b,\mu,\nu) \,\tilde H^{(3)N}_{q/q}(\mu),
\end{align}
with the renormalized matching coefficients
\begin{align}
    C_{qq}(\vect b, \mu,\nu) = &\, 1+\frac{\alpha_s}{2\pi}\left[C_F \left(  \ln\frac{\mu^2}{\mu_b^2} \left(\frac{3}{2} +2\ln\frac{\nu}{\omega_J}\right) + \frac{1}{N(N+1)}\right)-\gamma_{qq}(N)\ln \frac{\mu^2}{\mu_b^2} -2\gamma_{qq}^{(1)}(N)\right] ,  
    \label{eq: Cqq} \\
    C_{gq}( \vect b,\mu,\nu) = & \,\frac{\alpha_s}{2\pi}\left[-\gamma_{gq}(N)\ln \frac{\mu^2}{\mu_b^2} -2\gamma_{gq}^{(1)}(N)  +C_F\frac{1}{N+1}\right],
    \label{eq: Cgq} \\
    C_{gg}(\vect b,\mu,\nu) = & \,1+\frac{\alpha_s}{2\pi}\left[ C_A \ln\frac{\mu^2}{\mu_b^2} \left(\frac{\beta_0}{2C_A} +2\ln\frac{\nu}{\omega_J}\right)-\gamma_{gg}(N)\ln \frac{\mu^2}{\mu_b^2}-2\gamma_{gg}^{(1)}(N)\right],
    \label{eq: Cgg} \\
    C_{qg}(\vect b, \mu,\nu) = & \frac{\alpha_s}{2\pi}\left[-\gamma_{qg}(N)\ln \frac{\mu^2}{\mu_b^2} -2\gamma_{qg}^{(1)}(N) +T_F\frac{2}{(N+1)(N+2)} \right],
    \label{eq: Cqg} \\
    \delta C_{qq}( \vect b,\mu,\nu) = & 1+\frac{\alpha_s}{2\pi}\left[ C_F\ln\frac{\mu^2}{\mu_b^2} \left(\frac{3}{2} +2\ln\frac{\nu}{\omega_J}\right) -\Delta_T\gamma_{qq}(N)\ln \frac{\mu^2}{\mu_b^2} -2\Delta_T\gamma_{qq}^{(1)}(N)\right] . \label{eq: TCgq}
\end{align}
The $z^{N-1}$-weighted collinear FFs coincide with the Mellin moments
\begin{equation}
    \tilde{F}_{j/i}^N(\mu) = \int\d z \, z^{N-1} \,F_{j/i}(z,\mu)
\end{equation}
for $F\in\{D,\,\hat H^{(3)}\}$. Here $D_{j/i}(z,\mu)$ is the unpolarized collinear FF and $\hat H_{q/q}^{(3)}(z,\mu)$ is the first moment of the Collins function in the convention of Refs.~\cite{Yuan:2009dw,Kang:2015msa}. We neglect the contribution from two-variable twist-3 FFs to the transversely polarized OPEC jet function, consistent with the concurrent practice in available global analyses of the Collins function parametrization extractions.  The anomalous dimensions are given by
\begin{align}
    \gamma_{qq}(N)& = C_F\left[\frac{3N^2-N-2}{2N(N+1)}-2\left( \gamma_E+\psi(N)\right) \right], \\
    \gamma_{gq}(N)& = C_F\left[\frac{N^2+N+2}{N(N^2-1)} \right] , \\
    \gamma_{gg}(N)& = 2C_A\left[\frac{11}{12}-\frac{2}{N}+\frac{1}{(N+1)(N+2)}- \gamma_E-\psi(N-1) \right]-\frac{2}{3}n_f T_F, \\
    \gamma_{qg}(N)& = T_F\left[\frac{N^2+N+2}{N(N+1)(N+2)} \right] , \\
    \Delta_T\gamma_{qq}(N)& = C_F\left[\frac{3}{2}-2\left( \gamma_E+\psi(N+1)\right) \right], 
\end{align}
with $\gamma_{ij}^{(1)}(N)={\d}[\gamma_{ij}(N)]/{\d N}$, $\Delta_T\gamma_{qq}^{(1)}(N)={\d} [\Delta_T\gamma_{qq}(N)]/{\d N}$, $\gamma_E$ the Euler--Mascheroni constant and $\psi(N)$ the polygamma functions.
We note that the quark-initiated unpolarized and the transversely polarized OPEC jet functions share the same ultraviolet divergences. Their corresponding renormalization group (RG) equations and rapidity renormalization group (RRG) equations are given by
\begin{align}
    \mu\frac{\d}{\d\mu} \ln J_q^{U/T}(  \vect b, \mu, \nu) &= \gamma_{\mu,\, q}^{D/H}(\omega_J,\mu, \nu),
\\
    \mu\frac{\d}{\d\mu} \ln J_g^{U}(  \vect b, \mu, \nu) &= \gamma_{\mu,\, g}^{D}(\omega_J,\mu, \nu),
\\
\nu\frac{\d}{\d\nu} \ln  J_q^{U/T}( \vect b, \mu, \nu) &= \gamma_{\nu,\, q}^{D/H}(b, \mu), \\
\nu\frac{\d}{\d\nu} \ln  J_g^{U}(\vect b, \mu, \nu) &= \gamma_{\nu,\, g}^{D}(b, \mu),
\end{align}
with the one-loop anomalous dimensions
\begin{align}
    \gamma_{\mu, q}^{D/H}(\omega_J,\mu, \nu) =& \frac{\alpha_s}{\pi} C_F \left[2 \ln\left(\frac{\nu}{\omega_J}\right) + \frac{3}{2}\right],
\\
\gamma_{\mu, g}^{D}(\omega_J,\mu, \nu) =& \frac{\alpha_s}{\pi} C_A \left[2 \ln\left(\frac{\nu}{\omega_J}\right) + \frac{\beta_0}{2C_A}\right],\\
\gamma_{\nu, q}^{D/H}(b,\mu) = & \frac{\alpha_s}{\pi} C_F \ln\left(\frac{\mu^2}{\mu_b^2}\right), \\
\gamma_{\nu, g}^{D}(b,\mu) = & \frac{\alpha_s}{\pi} C_A \ln\left(\frac{\mu^2}{\mu_b^2}\right).
\end{align}

At NLL, the hard matching coefficients in  \cref{eq: OPEC JU,eq: OPEC JT} reduce to $\delta_{qq'}\,\delta(1-z)$, corresponding to the limit with no out-of-jet energy loss, where the jet carries the full energy of the initiating quark ($z=E_J/Q=1$). Consequently, using the RG and RRG equations given above, the OPEC FJFs with a linear energy weight ($z_h^1$) reduce to the expressions in \cref{eq:FJF_ope,eq:FJF_ope_Collins} of the main text, where we have ignored the jet radius $R$ resummation and the NLL perturbative Sudakov factor of the quark jet is given by
\begin{equation}
    S_{\rm pert}(Q,b)=\int^Q_{\mu_b} \frac{\d \mu'}{\mu'} \left[ A\,\ln\left(\frac{Q^2}{\mu'^2} \right)+ B \right],
\end{equation}
with $A = \sum_{n=1} A^{(n)}(\alpha_s/\pi)^n$, $B = \sum_{n=1} B^{(n)}(\alpha_s/\pi)^n$ and
\begin{align}
    &A^{(1)}=C_F, \qquad \qquad \qquad  B^{(1)}=-\frac{3}{2}C_F, \nonumber \\
    &A^{(2)}=\frac{C_F}{2} \left[ C_A\left(\frac{67}{18}-\frac{\pi^2}{6} \right)-\frac{10}{9}T_F n_f\right].
\end{align}

\subsection*{Scale uncertainties in SSA predictions}

In this section, we present the theoretical scale uncertainties associated with the SSA evaluations in both approaches discussed in the main text. For the TMD evolution framework, we vary both the OPE scale $\mu_i$ and the hard scale $\mu=Q$ independently by a factor of two around their canonical values ($\mu_i=\mu_{b^*}$ and $\mu=Q=p_T$) to estimate the perturbative uncertainty. For the JAM3D-based approach, where only collinear DGLAP evolution is included, we vary the hard scale $\mu=Q=p_T$ in the same manner.

\begin{figure}[!htb]
    \centering
    \includegraphics[width=1\linewidth]{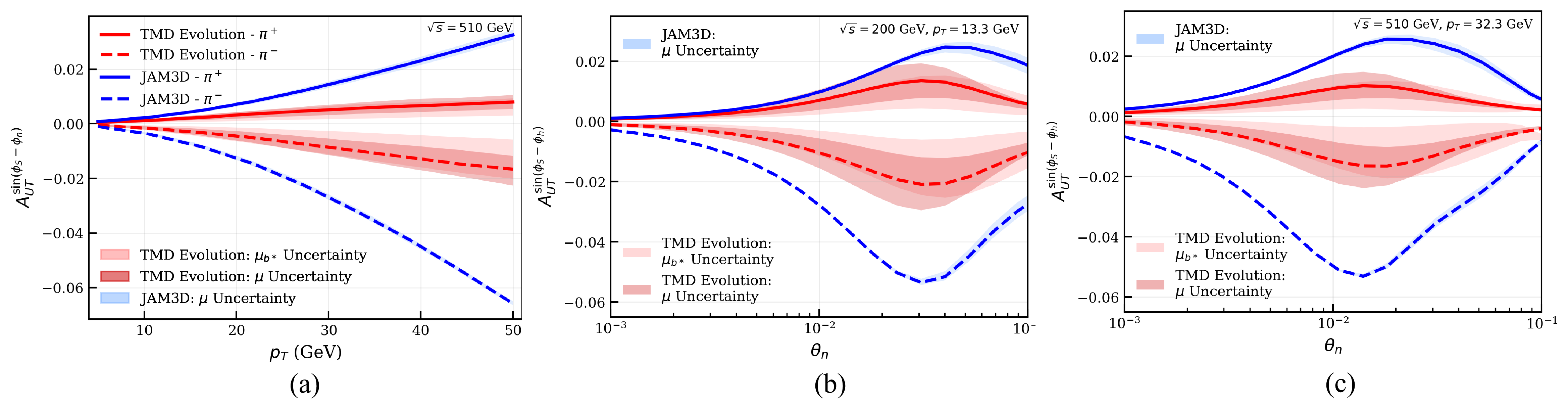}
    \caption{Theoretical uncertainties from scale variation for the OPEC SSA $A_{UT}^{\sin(\phi_s - \phi_n)}$ in $\pi^{\pm}$ production. 
    (a) Dependence on jet transverse momentum $p_T$ at $\sqrt{s}=510~\text{GeV}$. 
    (b) and (c) Dependence on the polar angle $\theta_n$ at 
    $\sqrt{s}=200$ and $510~\text{GeV}$, respectively. 
    The shaded bands represent variations of $\mu_{b^*}$ and $p_T$ by a factor of two.}
    \label{fig: Uncertainties}
\end{figure}

\Cref{fig: Uncertainties} shows the resulting scale bands for the OPEC SSA $A_{UT}^{\sin(\phi_s - \phi_n)}$ in charged pion ($\pi^{\pm}$) production. Panel (a) displays the asymmetry as a function of the jet transverse momentum $p_T$ at $\sqrt{s} = 510~\text{GeV}$, while panels (b) and (c) show its dependence on the energy flow polar angle $\theta_n$ for $\sqrt{s} = 200$ and $510~\text{GeV}$, respectively.

We find that the uncertainties arising from the scale variations in the NLL TMD evolution are comparable in magnitude to those from the parametrization of the nonperturbative collinear inputs. In contrast, the JAM3D results exhibit significantly smaller scale sensitivity, reflecting the absence of resummed double logarithms in the DGLAP-only evolution framework.

\end{document}